\documentclass[review]{elsarticle}

\usepackage{hyperref}
\usepackage{amsmath}

\journal{Journal of \LaTeX\ Templates}

\usepackage{graphicx}
\graphicspath{ {./NPApics/} }
\date{}

\begin{document}

\begin{frontmatter}

\title{Fusion hindrance of  $^{48}$Ti + $^{138}$Ba reaction}


\author[a]{K.K. Rajesh}
\ead{rajeshmlpm@yahoo.com}

\author[a]{M.M. Mustahfa}

\address[a]{Department of physics, University of Calicut, Kerala, India}

\begin{abstract}
We have already measured the ER excitation function of $^{48}$Ti + $^{138}$Ba reaction around the Coulomb barrier. While a detailed report of measurement of ER excitation function of $^{48}$Ti + $^{138}$Ba is available in literature, the present paper compares it with other channels forming the same compound nucleus. The experiment was performed at IUAC, New Delhi for lab energies from 189.3 to 234.4 MeV, where the ER cross sections show an abrupt fall beyond 160.03 MeV. The decrease in ER cross section is attributed to non compound nuclear reaction such as quasi- fission (QF). A comparison with statistical model code PACE 4 also shows significant deviation of the experimental data from theoretical predictions. When this reaction is compared with $^{86}$Kr + $^{100}$Mo and $^{64}$Ni + $^{122}$Sn forming the same compound nucleus $^{186}Pt^{*}$, the symmetric systems show more ERs than the asymmetric one and it is against the general trend. Preliminary study predict the influence of deformation and isospin asymmetry in the abrupt fall of ER cross section than that of the symmetric systems. A detailed theoretical analysis of the systems are in progress and will be published shortly.
 
\end{abstract}


\end{frontmatter}

\section{Introduction}

Research into the formation of heavy nuclei has contributed important informations into the reaction dynamics of fusion process. Overcoming the Coulomb barrier alone does not guarantee the formation of a compound nucleus (CN) in the nuclear reactions involving heavy reaction partners. The factors influencing the dynamical evolution of composite system, after the contact of two heavy nuclei, is not yet fully understood.

The formation of heavy nuclei involves three distinct steps with each step has its own influence in the formation of final product. First, the interacting nuclei overcome the Coulomb barrier which is repulsive in nature and approach close enough so that attractive nuclear force come into play. This results in the capture of projectile inside the Coulomb barrier. Second, the composite system or the di-nuclear system (DNS) undergo shape evolution towards a compact mononuclear shape or CN Finally, the composite system (CN) so formed should survive fission to form evaporation residue (ER)

Among these, the second step is considered to be both most complex and least understood stage in the formation of a heavy nucleus \cite{1}. Mathematically, the E.R. cross section is given by,
\begin{equation}
\sigma_{ER}= \sum_{J=0}^{\infty} \sigma_{J}(E_{c.m.},J) P_{CN}(E^{*},J)W_{sur}(E^{*},J)
\end{equation}
where $\sigma_{J}(E_{c.m.},J)$is the capture cross section as a function of center-of-mass energy $E_{c.m.}$ and angular momentum J$\hbar$, $P_{CN}(E^{*},J)$ is the probability that the system reaches the equilibrium configuration as a function of the excitation energy
$E^{∗}$ and J , and $W_{sur}(E^{*},J)$ is the probability that the system survives statistical fission decay through sequential particle evaporation, thus eventually forming a heavy evaporation residue.

Here $P_{CN}$ represent the second stage of heavy ion fusion. If $P_{CN}<<1$ it will result in non compound nuclear reaction such as quasi-fission (QF) and if $P_{CN}\simeq1$, it will result in CN formation. The competition between fusion-fission and QF during the evolution of the composite system is a very complex process. This competition is found to be influenced by entrance channel properties such as charge product of the entrance channel \cite{2,3,4,5}, deformation alignment \cite{6,7,8,9,10} , magicity \cite{11} and asymmetry of the projectile and target $\frac{N}{Z}$ ratios \cite{12}. A detailed study of all these factors will throw more and more light into the complex process of evolution of the composite system which ultimately results either in the formation of s compact CN or in QF.

Keeping all these objectives in mind, we present the comparison of ER excitation function measurement of the reaction $^{48}$Ti + $^{138}$Ba \cite{12.a}  with  more symmetric systems,$^{86}$Kr + $^{100}$Mo and $^{64}$Ni + $^{122}$Sn, which are also forming the same CN to get an insight into the various factors influencing the probability of CN formation.
\begin{figure}[h]

\centering

\includegraphics[width=8cm, height=6cm] {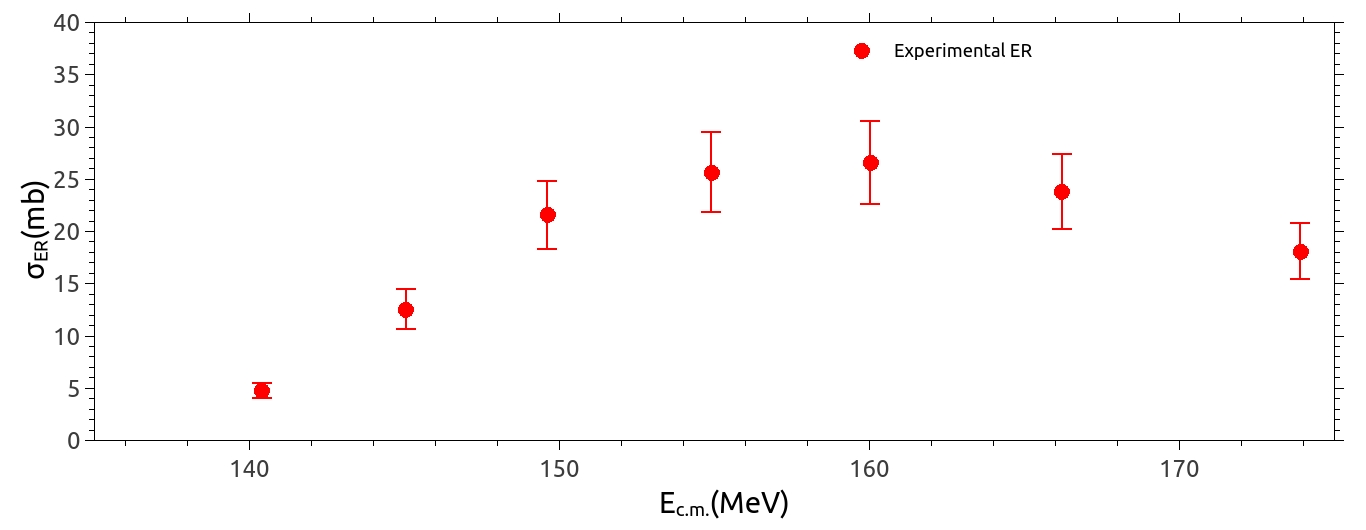}

\caption{\small{Figure 1. Total ER cross section as a function of centre of mass energy. The ER cross section falls significantly at higher beam energies.}}
\end{figure}
\section{ER excitation function of $^{48}$Ti + $^{138}$Ba}
\qquad The total ER cross section ($\sigma_{ER}$) of $^{48}$Ti + $^{138}$Ba \cite{12.a} obtained as a function of centre of mass energy ($E_{cm}$) is reproduced in Fig.1. The over all error in the calculated cross section is $\sim 15\%$ .The measured excitation function shows a decreasing trend at higher beam energies \cite{12.a}. 
One may assume that this decrease in ER cross section may be due to increased fission competition at larger angular momenta. But a high fission barrier of 16.3 MeV stems the system from decaying through fission channel. This suggest that at higher excitation energies the probability of CN formation is severely hindered. 

 The measured ER cross sections for each centre of mass energy along with corresponding excitation energy is shown in table 1. A theoretical comparison of experimental result with PACE4 statistical model code is shown in Fig.2. Statistical models assume complete fusion between projectile and target. So a deviation from statistical prediction suggest hindrance to ER formation due to some non CN process.

\begin{table*}[ht] 
\caption{\small {Measured ER cross section as a function of centre of mass energy and corresponding excitation energy .} }

\label{table:1}
\begin{center}
\begin{tabular}{c c c c} 
\hline\hline
$E_{CM}$(MeV) & $E^*$(MeV) & $\sigma_{ER}$(mb) \\
\hline
140.4 & 41.51 & 4.8 $\pm$ 0.7 \\ 
 145.04 & 46.15 & 12.6 $\pm$ 1.9\\ 
 149.6 & 50.71 & 21.6 $\pm$ 3.2 \\
 154.9 & 56.01 & 25.7 $\pm$ 3.9 \\
 160.03 & 61.14 & 26.6 $\pm$ 4.0 \\
 166.2 & 67.31 & 23.8 $\pm$ 3.6 \\
173.9 & 75.01 & 18.1 $\pm$ 2.7 \\
\hline\hline
\end{tabular}
\end{center}
\end{table*}

\begin{figure}[h]

\centering
\includegraphics[width=8cm, height=6cm] {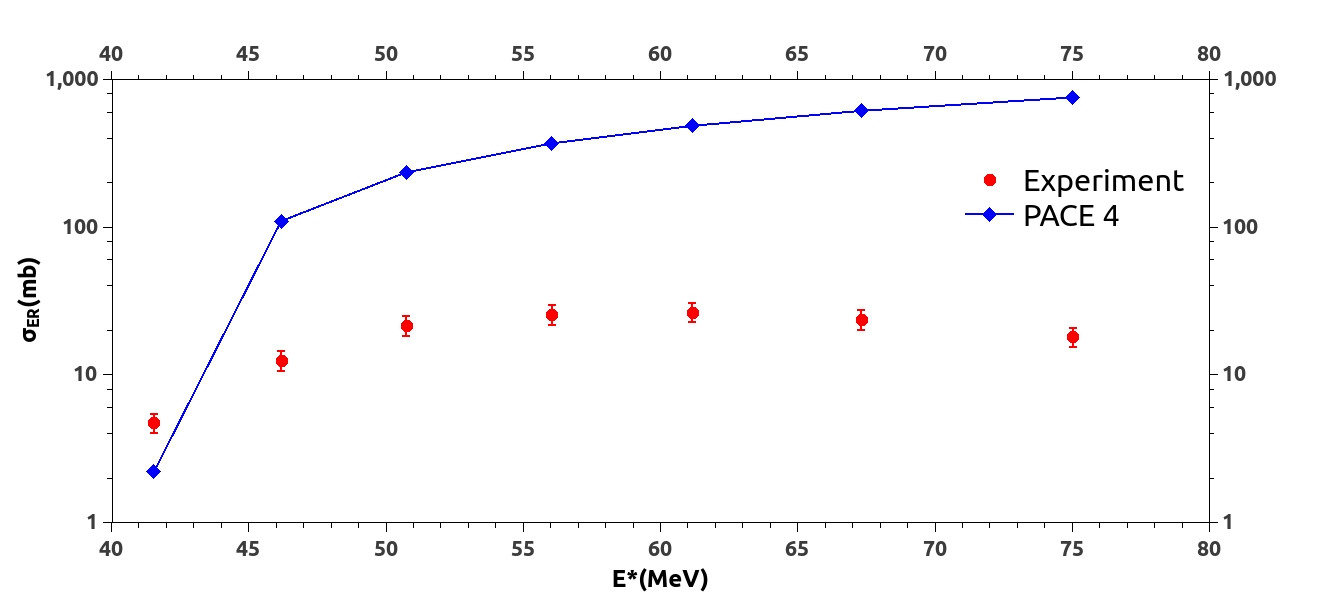}

\caption{\small {Figure 2.The experimental ER cross section is compared with PACE 4 statistical calculations. It is found that the experimental cross section deviate significantly from the statistical predictions.} }
\end{figure}

The survival probability against fission is shown in Fig.3. The experimental survival probability is obtained as the ratio of experimental ER cross section and ccfull code predicted capture cross section. Capture cross section is the sum of fusion cross section and cross section for non compound nuclear reactions ($\sigma_{Cap}= \sigma_{fus} + \sigma_{NCN}$). A high fission barrier of 16.3 MeV prevent the system from decaying through fission. So fusion cross section correspond to ER cross section (since $\sigma_{fus} = \sigma_{fission} +\sigma_{ER}$), which has a decreasing trend. Thus the deviation of this ratio from unity suggest a lesser probability for fusion, enhancing the chances of non CN reaction. It is found that the survival probability of $^{48}$Ti  + $^{138}$Ba has a deep fall which suggest a lesser fusion probability for this system.

\begin{figure}[h]

\centering  
\includegraphics[width=8cm, height=6cm] {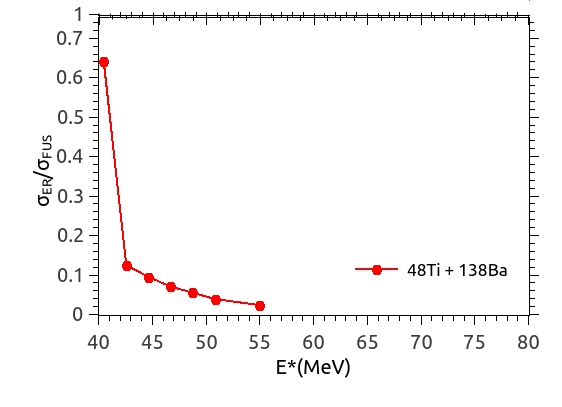}

\caption{\small {Figure 3.The ratio of ER cross section to fusion cross section gives survival probability against fission.} }
\end{figure}

\section{Comparison with Symmetric systems}
An investigation into the reasons for the fusion hindrance of the present experiment $^{48}$Ti  + $^{138}$Ba is performed by comparing it with other reactions, $^{86}$Kr + $^{100}$Mo  \cite{14} and $^{64}$Ni+$^{122}$Sn \cite{14a}. The three systems are identical in terms of entrance channel shell closure and compound nucleus formed. Both $^{48}$Ti  + $^{138}$Ba and $^{86}$Kr + $^{100}$Mo have single neutron magicity in the entrance channel, while $^{64}$Ni+$^{122}$Sn shows a single proton magicity in the entrance channel.
The various parameters of the two reactions is given in Table 2 and a comparative study of the reactions is shown in Figure 4.

\begin{figure}[h]

\centering
\includegraphics[width=8cm, height=6cm] {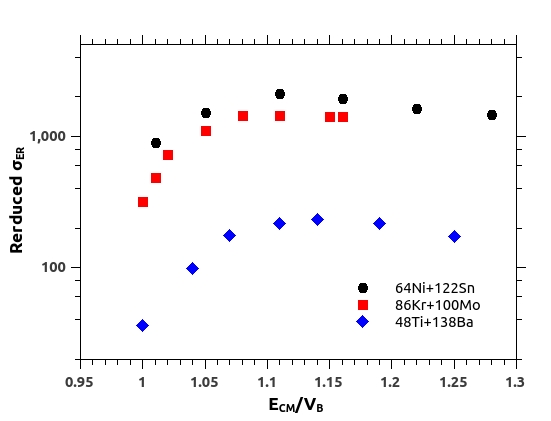}

\caption{\small {Figure 4.A comparative study of the two reactions show a reduced fusion probability for $^{48}$Ti + $^{138}$Ba reaction.} }
\end{figure}

\begin{table*}[ht]
\caption{Various parameters of the interacting nuclei through two different channels forming same compound nucleus Pt$^{186*}$.}
\centering
\begin{tabular}{p{0.16\linewidth}p{0.10\linewidth}p{0.07\linewidth}p{0.06\linewidth}p{0.05\linewidth}p{0.09\linewidth}p{0.09\linewidth}p{0.10\linewidth}}
\hline\hline
Reaction & Nucleus & $\beta_{2}$& N & Z & $\frac{N}{Z}$&$\Delta \frac{N}{Z}$ &  Magicity\\
\hline
& \\
$^{48}$Ti+$^{138}$Ba & Ti & 0.28 & 26 &22 & 1.18&  &-  \\
($Z_{p}Z_{t}$=1232)& & & & & &0.28 &  \\
 & Ba & 0.09 & 82 & 56 & 1.46&  & N=82 \\
\hline

$^{86}$Kr+$^{100}$Mo & Kr & 0.13 & 50 & 36 & 1.38& & N=50  \\
($Z_{p}Z_{t}$=1512) & & & & & & 0 & \\
& Mo & 0.16 & 58 & 42 & 1.38 &  & -  \\
\hline

$^{64}$Ni+$^{122}$Sn & Ni & 0.16 & 36 & 28 & 1.28 & &-   \\
($Z_{p}Z_{t}$=1400) & & & & & & 0.16 & \\
& Sn & 0.10 & 72 & 50 & 1.44 &  &   Z=50\\
\hline\hline

\end{tabular}
\end{table*}

All the three curves have a similar trend of decreasing ER cross section with increasing beam energy above the Coulomb barrier. Among the two reactions plotted above,$^{48}$Ti + $^{138}$Ba is the most asymmetric reaction. As a result the Coulomb repulsion between the target and projectile will be minimum and comparatively larger possibility for CN formation. This should have been resulted with a better ER production for $^{48}$Ti + $^{138}$Ba than the other two reactions. But an opposite trend has been observed in the comparative plot and it is not expected in general. 

\begin{figure}[h]

\centering

\includegraphics[width=8cm, height=6cm] {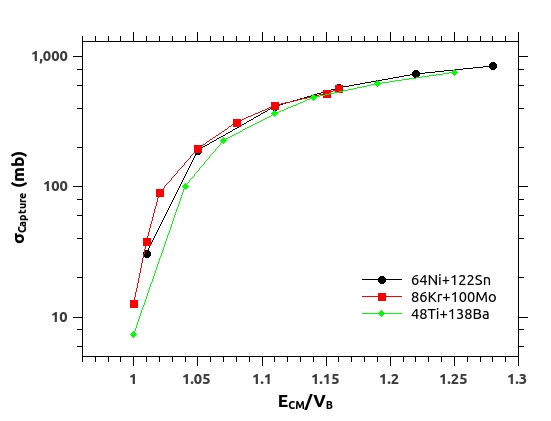}

\caption{\small {Figure 5.The capture cross section of all the systems forming $^{186}Pt^{*}$.} }
\end{figure}
The capture cross section of the three systems is shown in Fig.5. The capture cross section is calculated using ccfull code. The various parameters used in the ccfull code is shown in table 3. It is seen that the capture cross section for the formation of a compact CN is almost similar for all the systems. A higher fission barrier (16.3 MeV) stems the system from decaying through fission channel. This demands an increasing ER cross section with higher beam energy. But a fall in ER cross section point towards a lesser probability of CN formation due to non CN reaction such as QF. This suggest that QF events are dominated in the asymmetric reaction $^{48}$Ti + $^{138}$Ba than that in symmetric reactions which is not a general trend.
A discussion into the factors influencing the fusion hindrance of $^{48}$Ti + $^{138}$Ba is presented in the following sections.

\begin{table}[ht]
\caption{\small{{Table 3.The potential depth ($V_{0}$), radius ($a_{0}$) and surface diffuseness ($r_{0}$)parameters used in the CCFULL calculation of $^{48}$Ti+$^{138}$Ba, $^{86}$Kr+$^{100}$Mo and $^{64}$Ni+$^{122}$Sn reactions.} }}

\begin{center}
\setlength{\tabcolsep}{12pt}
\begin{tabular}{|c| c| c| c|} 
\hline\hline

Reaction& V$_{0}$  & a$_{0}$ & r$_{0} $ \\
& (MeV)&(fm) &(fm) \\
\hline

$^{48}$Ti+$^{138}$Ba & 79.85 & 0.685 & 1.18 \\ 
 $^{86}$Kr+$^{100}$Mo & 83.5 & 0.692 & 1.18 \\
 $^{64}$Ni+$^{122}$Sn & 82.4 & 0.69 & 1.18 \\ 

\hline\hline
\end{tabular}
\end{center}
\end{table}

\subsection{Influence of deformation}
It has been reported that static deformation of the heavy reaction partner affects nuclear collisions \cite{15}. This is because the capture barrier height depends on the relative orientation of the projectile and the deformation axis of statistically deformed heavy reaction partner \cite{16}. Among the three reactions being compared, $^{48}$Ti + $^{138}$Ba has largest deformation parameter with $\beta=0.28$ for $^{48}Ti$. It has been widely shown that the collisions with tips of the prolate deformed target nucleus result in Q.F. \cite{17,18}. This is due to the formation of an elongated composite system which reseparates before the equilibration in all degrees of freedom. The reason for this phenomenon can be explained with the help of potential well formation. For reactions with a deformed nuclei, potential well for side to side orientation is deeper and wider than that for tip to tip collisions. Thus, the number of partial waves leading to capture and fusion are comparatively smaller for tip to tip collisions. In other words, only collisions with side to side orientation will lead to fusion, leaving all other orientation probably leading to QF.

A study on the role of orientation angles of symmetry axis of deformed projectile and target nucleus relative to beam direction in the fusion and capture process of heavy ion collision shows that in the tip to tip collisions of nuclei an increase in beam energy does not lead to increase in the fusion cross section \cite{18}. This suggest an increased competition from QF mechanism at higher excitation energies in the present study. The suppression of ER in $^{48}$Ti+$^{138}$Ba, $^{86}$Kr+$^{100}$Mo and $^{64}$Ni+$^{122}$Sn reactions suggest the presence of QF in all the reactions. Being more asymmetric reaction, $^{48}$Ti + $^{138}$Ba is supposed to have more fusion probability. But the significant suppression noticed in this reaction can be attributed to deformation effect in its entrance channel.

\subsection{Influence of isospin asymmetry}  

Some recent studies have shown that the difference in the $\frac{N}{Z}$ ratio of the reaction partners can influence the reaction dynamics\cite{19,20}.The study of dependence of $\frac{N}{Z}$ asymmetry or isospin asymmetry of projectile and target on the reaction dynamics is relatively a new entrant in the field of heavy ion induced nuclear reactions. The isospin asymmetry is quantified by the difference between $\frac{N}{Z}$ ratios of the initial colliding nuclei\cite{19}.This difference is represented as $\Delta \frac{N}{Z}$ in this paper.  All the systems have single entrance channel magicity or shell closure and almost similar charge product. Despite having shell closure in the entrance channel, identical charge product of the reaction partners and forming same CN, there is a significant variation on the reaction outcomes.

From Table 2, it can be seen that the value of $\Delta \frac{N}{Z}$ is larger with a value of 0.28 for $^{48}$Ti + $^{138}$Ba  reaction. The reaction with a larger value of $\Delta \frac{N}{Z}$ shows significant suppression in the ER cross section. It has been reported that magic numbers in the entrance channel with large isospin asymmetry $(\Delta \frac{N}{Z})$ increase QF \cite{19,20} and hence it is reflected as a reduction in the ER cross sections. The explanation for this phenomenon based on TDHF calculation is that, during the interaction between collision partners with large isospin values a rapid equilibration takes place in the early stage of the reaction  which modifies the identities of the colliding partners. Prior to the capture process, protons and neutrons are exchanged between nuclei to equalise $\frac{N}{Z}$ ratio. Thus matching of $\frac{N}{Z}$ ratio is an important condition in the enhancement of fusion reactions of magic nuclei \cite{21}.
We interpret this $\frac{N}{Z}$ mismatch between the reaction partners as the underlying reason for the reduced cross section of $^{48}$Ti + $^{138}$Ba reaction, even though it is the asymmetric system among the three reactions under comparison.

\section{Conclusion} 
To summarise, the $^{48}$Ti + $^{138}$Ba reaction shows an abrupt fall in the ER excitation function measurement for an excitation energy beyond 61MeV. This is attributed to non compund nuclear reaction called QF. A  comparison with more symmetric systems with shell closure shows the influence of deformation and isospin asymmetry in the fusion hindrance of present system. A detailed theoretical analysis of $^{48}$Ti+$^{138}$Ba, $^{86}$Kr+$^{100}$Mo and $^{64}$Ni+$^{122}$Sn systems are in progress. 
Preliminary study suggest that deformation and isospin asymmetry have a dominant role in the abrupt fall of ER cross section than that of the symmetric systems.Since a very less number of studies have been conducted in the heavy ion induced reaction with entrance channel shell closure, more such studies have a wide scope of revealing the underlying phenomenon that hinders fusion.

\section*{Acknowledgement} 

One of the authors (KKR) is grateful to N. Madhavan, S. Nath, S.R. Abhilash, J. Gehlot  and M. M. Hosamani for the ER excitation measurement at IUAC, New Delhi. Thanks are due to E. Prasad and Jhilam Sadhukhan for their help and cooperation.


\section*{Bibliography}
\begin{thebibliography}{100} 

\bibitem{1} V. I. Zagrebaev and W. Greiner, J. Phys. G: Nucl. Part. Phys. 31, 825 (2005).
\bibitem{2} C. C. Sahm, H. G. Clerc, K.-H. Schmidt, W. Reisdorf,P. Armbruster, F. P. Hessberger, J. G. Keller, G. Munzenberg and D. Vermeulen, Z. Phys. A 319, 113 (1984).
\bibitem{3}A. C. Berriman, D. J. Hinde, M. Dasgupta, C. R. Morton,R. D. Butt, and J. O. Newton, Nature (London) 413, 144(2001).
\bibitem{4}D. J. Hinde, M. Dasgupta, and A. Mukherjee, Phys. Rev. Lett. 89, 282701 (2002).
\bibitem{5}D. J. Hinde and M. Dasgupta, Phys. Lett. B 622, 23 (2005)
\bibitem{6}D. J. Hinde, M. Dasgupta, J. R. Leigh, J. P. Lestone, J. C. Mein,C. R. Morton, J. O. Newton, and H. Timmers, Phys. Rev. Lett.74, 1295 (1995).
\bibitem{7}D. J. Hinde, M. Dasgupta, J. R. Leigh, J. C. Mein, C. R.Morton, J. O. Newton, and H. Timmers, Phys. Rev. C 53, 1290(1996).
\bibitem{8}G. N. Knyazheva, E. M. Kozulin, R. N. Sagaidak, A. Y. Chizhov,M. G. Itkis, N. A. Kondratiev, V. M. Voskressensky, A. M.Stefanini, B. R. Behera, L. Corradi, E. Fioretto, A. Gadea, A.Latina, S. Szilner, M. Trotta, S. Beghini, G. Montagnoli, F.Scarlassara, F. Haas, N. Rowley, P. R. S. Gomes, and A. S. d.Toledo, Phys. Rev. C 75, 064602 (2007).
\bibitem{9}K. Nishio, H. Ikezoe, S. Mitsuoka, I. Nishinaka, Y. Nagame, Y.
Watanabe, T. Ohtsuki, K. Hirose, and S. Hofmann, Phys. Rev.C 77, 064607 (2008).
\bibitem{10}K. Nishio, S. Mitsuoka, I. Nishinaka, H. Makii, Y. Wakabayashi,H. Ikezoe, K. Hirose, T. Ohtsuki, Y. Aritomo, and S. Hofmann,Phys. Rev. C 86, 034608 (2012).
\bibitem{11}G. Mohanto,D. J. Hinde, K. Banerjee, M. Dasgupta,D. Y. Jeung, C. Simenel,E. C. Simpson, A. Wakhle,E. Williams, I. P. Carter, K. J. Cook,  D. H. Luong, C. S. Palshetkar, and D. C. Rafferty, Phys. Rev.C 97, 0054603 (2018).
\bibitem{12}C. Simenel, D. J. Hinde, R. du Rietz, M. Dasgupta, M. Evers,C. J. Lin, D. H. Luong, and A. Wakhle, Phys. Lett. B 710, 607 (2012).
\bibitem{12.a} K. K. Rajesh, M. M. Musthafa, N. Madhavan, S. Nath, J. Gehlot, Jhilam Sadhukhan, P.Mohamed Aslam, P. T. Muhammed shan, E. Prasad, M. M. Hosamani, T. Varughese, Abhishek Yadav, Vijay R. Sharma, Vishal Srivastava, Md. Moin Shaikh, M. Shareef, A. Shamlath, and P. V. Laveen, Phys. Rev. C 100, 044611 (2019)

\bibitem{14} W. Reisdorf,F.P. Hessberger, K.D. Hildenbrand,G. Munzenberg,K.H.Schmidt, W.F.W.Schneider, and G. Wirth, Nuclear PhysicsA 444 (1985).
\bibitem{14a} W. S. Freeman,H. Ernst,D. F. Geesaman, W. Henning, T. J. Humanic,W. Kuhn,G. Rosner, J. P. Schiffer, B. Zeidman and F. W. Prosser, Phys. Rev. Lett. 50, 20 (1983). 
\bibitem{15}D.J. Hinde , M. Dasgupta , D.Y. Jeung , G. Mohanto , E. Prasad, C. Simenel , E.Williams , I.P. Carter ,K.J. Cook , Sunil Kalkal , D.C. Rafferty , E.C. Simpson , H.M. David , Ch.E. Düllmann , and J. Khuyagbaatar, EPJ Web of Conferences 163, 00023 (2017)
\bibitem{16}M. Dasgupta, D.J. Hinde, N. Rowley, A.M. Stefanini, Annu. Rev. Nucl. Part. Sci. 48, 401 (1998)
\bibitem{17}D.J. Hinde, C.R. Morton, M. Dasgupta, J.R. Leigh, J.C. Mein and H. Timmers, Nuclear Physics A 592 (1995) 271-289
\bibitem{18} Avazbek Nasirov, Akira Fukushima, Yuka Toyoshima, Yoshihiro Aritomo, Akhtam Muminov, Shuhrat Kalandarov and Ravshanbek Utamuratov,Nuclear Physics A 759 (2005) 342–369 

\bibitem{19}C. Simenel , D.J. Hinde , R. du Rietz  , M. Dasgupta , M. Evers  , C.J. Lin , D.H. Luong  and A. Wakhle, Physics Letters B 710 (2012) 607–611 
\bibitem{20} G. Mohanto, D. J. Hinde, K. Banerjee, M. Dasgupta, D. Y. Jeung, C. Simenel, E. C. Simpson, A. Wakhle, E. Williams, I. P. Carter, K. J. Cook, D. H. Luong, C. S. Palshetkar, and D. C. Rafferty, Phys. Rev. C 97,  (2018)054603
\bibitem{21} D.J. Hinde, R. du Rietz, E. Williams, C. Simenel, C.J. Lin , A. Wakhle , K.J. Cook , M.Dasgupta , M. Evers, and D.H. Luong, EPJ Web of Conferences 66 , (2014)03037 





\end{thebibliography}
\end{document}